\documentclass[sn-mathphys-num]{sn-jnl}

\newif\ifblind
\blindfalse

\usepackage{graphicx}
\usepackage{graphicx}%
\usepackage{multirow}%
\usepackage{amsmath,amssymb,amsfonts}%
\usepackage{amsthm}%
\usepackage{mathrsfs}%
\usepackage[title]{appendix}%
\usepackage{xcolor}%
\usepackage{textcomp}%
\usepackage{manyfoot}%
\usepackage{booktabs}%
\usepackage{algorithm}%
\usepackage{algorithmicx}%
\usepackage{algpseudocode}%
\usepackage{listings}%
\usepackage{subcaption}
\usepackage{longtable}
\usepackage{pdflscape}
\usepackage{hyperref}

\theoremstyle{thmstyleone}%
\theoremstyle{thmstyletwo}%

\theoremstyle{thmstylethree}%

\begin{document}

\title[Clustering Astronomical Orbital Synthetic Data]{Clustering  Astronomical Orbital Synthetic Data Using Advanced Feature Extraction and Dimensionality Reduction Techniques}

%
%
%
%
%
%




\ifblind
\else
  \author[1]{Eraldo Pereira Marinho}
  \author[1]{Nelson Callegari Junior}
  \author[1]{Fabricio Aparecido Breve}
  \author[1]{Caetano Mazzoni Ranieri}
  \affil[1]{Institute of Geosciences and Exact Sciences (IGCE), Unesp, Rio Claro, Brazil}
\fi


\abstract{
The dynamics of Saturn's satellite system offer a rich framework for studying orbital stability and resonance interactions. Traditional methods for analysing such systems, including Fourier analysis and stability metrics, struggle with the scale and complexity of modern datasets. This study introduces a machine learning-based pipeline for clustering $\sim22,300$ simulated satellite orbits, addressing these challenges with advanced feature extraction and dimensionality reduction techniques.
The key to this approach is using MiniRocket, which efficiently transforms 400 timesteps into a 9,996-dimensional feature space, capturing intricate temporal patterns. Additional automated feature extraction and dimensionality reduction techniques refine the data, enabling robust clustering analysis. This pipeline reveals stability regions, resonance structures, and other key behaviours in Saturn’s satellite system, providing new insights into their long-term dynamical evolution.
By integrating computational tools with traditional celestial mechanics techniques, this study offers a scalable and interpretable methodology for analysing large-scale orbital datasets and advancing the exploration of planetary dynamics.
}




\maketitle

\section{Introduction}\label{sec1}

{The temporal evolution of celestial bodies' orbits, particularly satellites' orbits around planets, is a complex process influenced by numerous dynamical factors. Departing with initial states close to the real orbit of a determined body, a sheer ensemble with typically tens of thousands of initial orbits is numerically integrated for very long periods. As a result, a massive amount of time series of representative orbital elements are generated for each clone orbit. A powerful tool often applied in dynamical astronomy is the construction of dynamical maps. This technique consists of plotting some stability criterion on a two-dimensional map, ready to furnish the main regions of the phase space with physical interest in terms of the long-term stability and interactions of clones of real satellites \cite{callyoko2010}. Clustering behaviours among simulated orbits offer a powerful framework for deciphering the dynamical structure of planetary systems. Researchers can identify stability zones, resonance structures, and evolutionary pathways by grouping orbits with similar properties furnished by the time series. For this task, traditional methods such as Fourier analysis and numerical stability metrics can be applied and are effective, though computationally expensive and struggle to scale with the large, high-dimensional datasets generated by modern simulations (e.g. \cite{callyoko2020}, \cite{calletal21}, \cite{callrodri23}).}

{This study presents a novel machine-learning-based pipeline designed to analyse and cluster the time series, effectively addressing the scalability and complexity limitations of traditional methods in celestial mechanics.}
At the core of this pipeline is MiniRocket~\cite{dempster2021mini}, a state-of-the-art method that performs feature extraction from time series using random convolutional kernels. In our case, MiniRocket transforms raw 400-step time series data into a high-dimensional feature space with 9,996 features. By utilising convolutional kernels with precisely tuned dilation and weights, MiniRocket efficiently captures intricate temporal patterns, making it particularly suited to the time series characteristics of orbital data. While recent advancements in time series clustering, such as random convolutional kernels, have demonstrated remarkable improvements in efficiency and scalability \cite{jorge2024time}, this study adopts traditional convolutional kernels for their interpretability and established effectiveness in capturing key dynamical features.
Additionally, TSFresh automates the extraction of interpretable features, enabling the identification of statistically significant patterns across the orbital dataset \cite{christ2018time}.

Dimensionality reduction techniques refine the high-dimensional features that MiniRocket,
and TSFresh extract. Linear methods, such as Principal Component Analysis (PCA), reduce dimensionality while preserving the most significant variances \cite{jolliffe2016principal}. Non-linear approaches, including t-SNE \cite{van2008visualizing}, UMAP \cite{mcinnes2018umap}, and autoencoders \cite{hinton2006reducing}, uncover complex, non‑linear relationships within the data. These combined methods enable a comprehensive and interpretable clustering analysis of the satellite assembly of orbits.


The remainder of this paper is structured as follows. Section~\ref{sec:2} provides a comprehensive review of related work in the context of time series clustering and orbital dynamics. In Section~\ref{sec:3}, we detail the proposed methodology, including the combined feature extraction procedures and dimensionality reduction strategies employed. Section~\ref{sec:4} addresses the hyperparameter optimisation process adopted throughout the study. A preliminary validation of the approach is presented in Section~\ref{sec:5}, through the clustering of time series generated by the classical simple pendulum. Section~\ref{sec:6} presents the application of the proposed clustering framework to a dataset composed of numerically integrated orbital time series corresponding to an ensemble of fictitious Saturnian satellites, inspired by discoveries from the Cassini-Huygens Planetary Mission. Finally, Section~\ref{sec:7} summarises the main conclusions and outlines potential directions for future work.

\section{Related work}\label{sec:2}

The analysis of orbital dynamics in celestial mechanics has traditionally relied on numerical simulations and stability metrics, such as those derived from Fourier analysis \cite{callyoko2010}. These methods have proven effective for understanding resonance structures and stability zones, particularly in planetary systems like Saturn's satellite system. However, their computational cost and inability to scale with large datasets pose significant challenges in the era of modern astronomical simulations.

Recent advances in machine learning have paved the way for more efficient and scalable approaches to analysing high-dimensional time series data. Feature extraction techniques like TSFresh \cite{christ2018time}.

The introduction of random convolutional kernels for time series analysis and clustering, as in \cite{jorge2024time}, represents a further step in improving scalability and efficiency. These methods leverage randomly initialized convolutional filters to extract meaningful patterns without the need for extensive training, offering competitive performance in clustering tasks. Similarly, MiniRocket \cite{dempster2021mini} has established itself as a state-of-the-art feature extractor, transforming raw time series data into a high-dimensional feature space that captures both local and global temporal dynamics with exceptional efficiency.

While these advancements have shown promise, gaps remain in applying them to large-scale astronomical datasets, such as those involving orbital dynamics. Many studies focus on general-purpose time series data or limited domains, leaving the integration of machine learning with classical astronomical methods underexplored. This study addresses these gaps by introducing a comprehensive pipeline that combines the scalability of MiniRocket with interpretable feature extraction methods and robust clustering techniques specifically tailored to the complex dynamical interactions in Saturn's satellite system.

\section{Methodology}\label{sec:3}

\begin{figure}[!ht]
    \centering
    \includegraphics[width=1.0\linewidth]{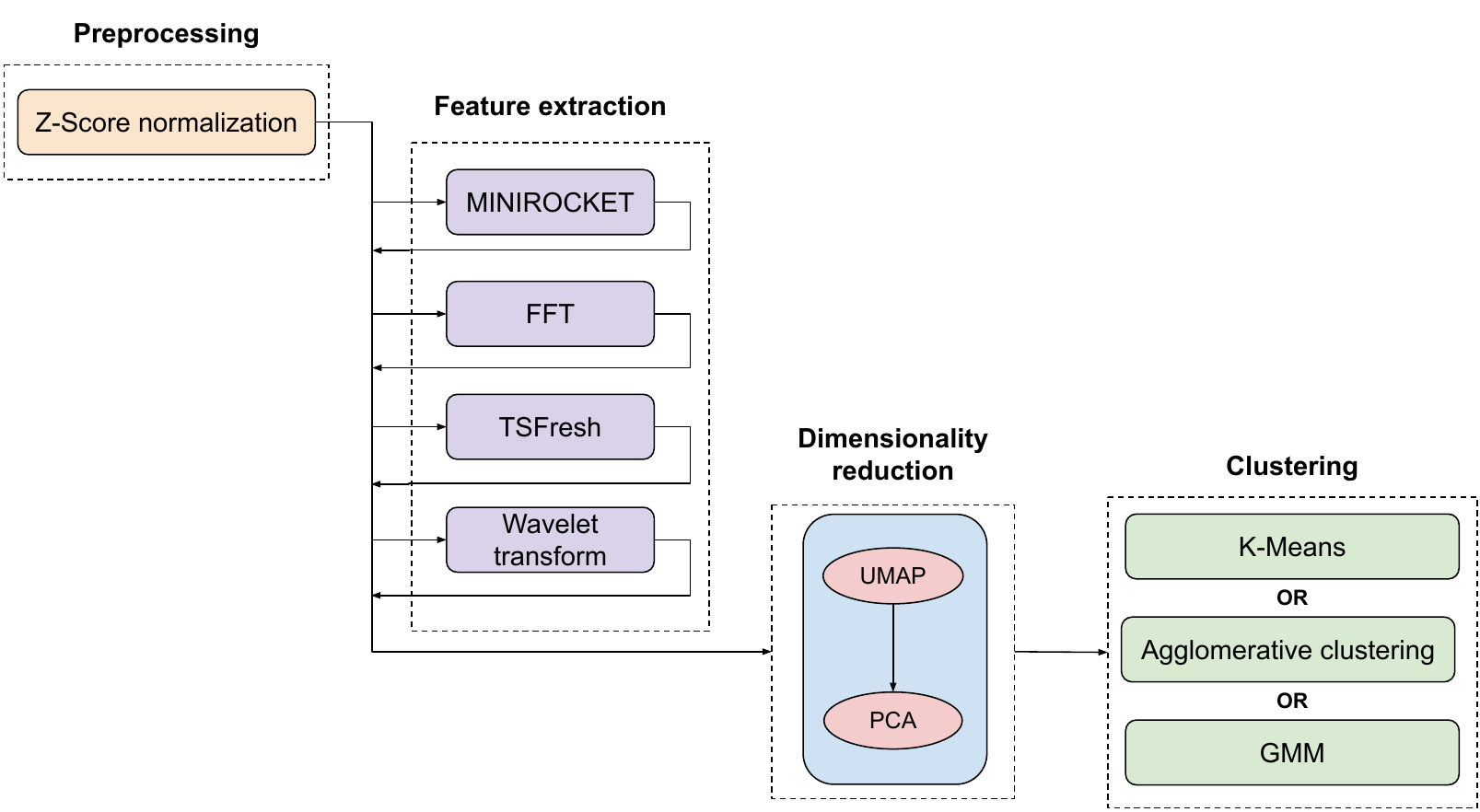}
    \caption{Proposed pipeline for clustering time-series data. The data shape at each stage is shown to improve reproducibility.
    Feature extraction modules are optional and are enabled or disabled to form different feature subsets (ablation study). Only the selected subset is concatenated to build the final feature vector; see Table~\ref{tab:feature_dimensions} for the evaluated configurations and best-performing pipelines.}
    \label{fig:pipeline}
\end{figure}

This study employs a machine learning-based pipeline to analyse and cluster the orbital dynamics of Saturn satellites. The data represent the orbits of small satellites in the Saturnian system, specifically focusing on the dynamics of Anthe (a small moon of Saturn) and its resonance interactions with Mimas (a mid-sized moon of Saturn). The numerical simulations model the orbital motion of the test satellites as they orbit Saturn, considering the gravitational influences of both the planet and its larger moons \cite{callyoko2020}.

The dataset $D$ consists of $22,288$ samples, each comprising a time series $d \in D$ with $400$ timesteps. Each sample corresponds to a pair of variables representing the initial state of the system. The variables contained in the time series are two angles $\varphi_1$ and $\varphi_2$, characterizing the system's capture in the so-called Corotation and Lindblad resonances, respectively, where motion remains confined around stable equilibria states. The definition of the angles and more details on the resonant dynamics is given in Section~\ref{sec:6}.

\subsection{Machine Learning Pipeline}
\label{subsec:pipeline}

The machine learning pipeline, shown in Figure~\ref{fig:pipeline}, was designed to integrate advanced feature extraction, dimensionality reduction, and clustering methods for data processing. The diagram is intended as a \emph{conceptual pipeline}, rather than a representation of parallel execution or data-flow concurrency. The subsequent subsections describe each step in the pipeline, including specific implementation details for reproducibility.

\subsubsection{Feature Extraction}

\textcolor{black}{
Before feature extraction, each time series is normalized using per-series z-score normalization: the mean and standard deviation are computed over the 400 timesteps of each orbit, and the orbit is standardized independently. No explicit detrending or windowing procedure is applied. This choice ensures that the feature extraction stage focuses on the temporal structure of the signal rather than absolute offsets or scale differences, and it also affects the interpretation of Euclidean and cosine distances after concatenating heterogeneous feature blocks.
}

\textcolor{black}{As shown in Figure 1, the proposed feature engineering pipeline is executed sequentially, with each module receiving the output of the previous step as input. Therefore, unlike parallel feature extraction schemes based on concatenation, each transformation progressively refines the time-series representation, and the resulting feature dimensionality evolves across stages according to the specific operation applied. Z-score standardization is applied to the input of each feature extraction module to normalize the signal scale. This preprocessing step does not alter the representation's dimensionality.
}

\textcolor{black}{
Each experiment activates a specific subset of extractors, while the remaining ones are disabled. This design implements an ablation strategy over feature representations. Conceptually, this stage is represented by a vertical feature bus, from which individual extractors can be selectively enabled. Each enabled extractor maps the normalized time series into a feature space of fixed dimensionality. When more than one extractor is active, the output of the previous block is fed to the input of the next one.
}

Four techniques were considered. These techniques were the MiniRocket method~\cite{dempster2021mini}, the Fast Fourier Transform (FFT) and DWT~\cite{rao2018digital}, and TSFresh~\cite{christ2018time}.
The resulting feature shapes at each step are detailed in Table~\ref{tab:feature_dimensions}.

MiniRocket is a method initially designed for time series classification. It transforms time series data into a feature space using convolutional kernels, enabling linear classifiers like ridge regression or logistic regression to achieve high accuracy with the generated feature vectors. Unlike its predecessor (i.e., Rocket \cite{dempster2020rocket}), MiniRocket employs a small, fixed set of carefully designed kernels, making it computationally efficient and up to 75 times faster on large datasets while maintaining comparable accuracy to state-of-the-art methods.

In this work, we adapt MiniRocket as a feature extractor for time series clustering. By processing its output feature vectors through dimensionality reduction techniques, we obtain a lower-dimensional representation suitable for clustering methods. This approach leverages MiniRocket's efficiency and effectiveness, extending its utility beyond classification to unsupervised tasks.

We have also applied other techniques to extract features from the time series and enhance the representation fed to the clustering methods.
The Fast Fourier Transform (FFT) breaks down a signal into its frequency components by efficiently computing the Discrete Fourier Transform (DFT) or its inverse, the Inverse Discrete Fourier Transform (IDFT). FFT converts a time-domain signal into its frequency-domain representation, revealing characteristics such as periodic patterns, dominant frequencies, and spectral energy distribution. This transformation is beneficial for analysing time series data, as it enables the extraction of frequency-based features that capture underlying patterns and trends.
Applying FFT produces a frequency vector with half the dimensionality of the input signal.

DWT decomposes a signal into scaled and shifted versions of a wave-like function known as a wavelet. Unlike the Fourier transform, which only provides frequency information and assumes that the signal is stationary, the DWT delivers a time-frequency representation. This representation reveals how different frequency components change over time. Wavelets offer a versatile framework for analysing real-world signals' timing and frequency characteristics, allowing for localized analysis across various scales.

Finally, TSFresh is a framework for automating feature extraction and selection in time series data. It combines signal processing and statistical techniques to uncover meaningful patterns. Central to TSFresh is the FRESH algorithm, which uses hypothesis testing to assess the relevance of extracted features for classification or regression tasks. This approach ensures that only significant features are retained, minimizing the risk of overfitting and improving model generalization. The framework offers 63 feature calculators that generate a total of 794 features, covering metrics such as distribution, entropy, and stationarity.

The dimensionality of the resulting feature vectors depends on the last feature extractor applied in the pipeline, following the dimensions shown in Table~\ref{tab:feature_dimensions}. Only the best-performing configurations are reported in the benchmark results (Table~\ref{tab:compact_pipeline_benchmarks}).

Table~\ref{tab:compact_pipeline_benchmarks} summarises a selection of the most relevant pipeline configurations, ranked according to the Silhouette score, followed by the Davies–Bouldin (DB) and Calinski–Harabasz (CH) indices. Each result corresponds to a pipeline whose parameters were optimised through an exhaustive grid search over the hyperparameter space. For each configuration, the table reports the feature extraction methods used, the clustering algorithm, the dimensionality reduction parameters (PCA and UMAP), the achieved clustering scores, and the sample distribution across the resulting clusters.

\begin{table}[!ht]
    \centering
    \caption{Feature dimensions at each extraction step.}
    \label{tab:feature_dimensions}
    \begin{tabular}{|l|c|}
        \hline
        \textbf{Feature Extraction Method} & \textbf{Feature Dimensions} \\
        \hline
        MiniRocket & (22,288, 9,996) \\
        FFT & (22,288, 201) \\
        Wavelet Transform & (22,288, 401) \\
        TSFresh & (22,288, 794) \\
        \textbf{Combined Features} & \textbf{(22,288, 11,375)} \\
        \hline
    \end{tabular}
\end{table}

\begin{landscape}
\begin{longtable}{llrrrrrl}
\caption{Pipeline Benchmark Table}
\label{tab:compact_pipeline_benchmarks}\\
\toprule
                    Features &        Method &  PCA &  UMAP &  Silhouette &     DB &          CH &                            Clusters \\
\midrule
\endfirsthead
\caption[]{Compact Benchmark Table (Landscape + Shrunk)} \\
\toprule
                    Features &        Method &  PCA &  UMAP &  Silhouette &     DB &          CH &                            Clusters \\
\midrule
\endhead
\midrule
\multicolumn{8}{r}{{Continued on next page}} \\
\midrule
\endfoot

\bottomrule
\endlastfoot
minirocket, wavelet, tsfresh &       K-means &    2 &    55 & 0.6830 & 0.4176 & 115173.8695 &  C0:9671, C1:4452, C2:2528, C3:5637 \\ \\
         minirocket, tsfresh & Agglomerative &    3 &    50 & 0.6830 & 0.4424 &  52505.8749 & C0:7847, C1:10152, C2:2527, C3:1762 \\ \\
    minirocket, fft, tsfresh &           GMM &    2 &    90 & 0.6810 & 0.4612 &  95632.9438 &  C0:9316, C1:5582, C2:4893, C3:2497 \\ \\
         minirocket, tsfresh &       K-means &    2 &    20 & 0.6809 & 0.4338 & 109598.8835 &  C0:5618, C1:9647, C2:2528, C3:4495 \\ \\
    minirocket, fft, tsfresh &       K-means &    2 &    60 & 0.6722 & 0.4334 & 106145.4625 &  C0:9685, C1:5625, C2:2525, C3:4453 \\ \\
minirocket, wavelet, tsfresh &           GMM &    2 &    50 & 0.6650 & 0.4308 & 104599.3775 &  C0:6254, C1:2528, C2:9208, C3:4298 \\ \\
\end{longtable}
\end{landscape}

\subsubsection{Dimensionality Reduction}

Dimensionality reduction (\textbf{DR}) is the next step in the pipeline. The feature vectors generated by the techniques discussed in the previous subsection are often high-dimensional, which complicates their application in clustering methods. This study utilised two complementary techniques to address this issue: Principal Component Analysis (PCA) for linear DR and Uniform Manifold Approximation and Projection (UMAP) for non-linear DR.

\paragraph{Principal Component Analysis (PCA)}

PCA reduces the dimensionality of data by projecting it onto a set of orthogonal axes that maximize variance. It is particularly effective for identifying the directions of the highest variance in a dataset, which often represent the most informative aspects of the data. Given a dataset with \( n \) samples and \( d \) features, represented as a two-dimensional matrix \( \mathbf{X} \in \mathbb{R}^{n \times d} \), it begins by centring the data to have zero mean and computing the covariance matrix $\mathbf{C}$, as shown in Equation~\ref{eq:covariance}.

\begin{equation}
    \label{eq:covariance}
    \mathbf{C} = \frac{1}{n} \mathbf{X}^\top \mathbf{X}
\end{equation}

Next, the eigenvalues \( \lambda_i \) and eigenvectors \( \mathbf{v}_i \) of \( \mathbf{C} \) are computed to satisfy Equation~\ref{eq:eigenvalues}, where \( \lambda_i \) indicates the amount of variance explained by the corresponding principal component \( \mathbf{v}_i \).

\begin{equation}
    \label{eq:eigenvalues}
    \mathbf{C} \mathbf{v}_i = \lambda_i \mathbf{v}_i
\end{equation}

The data is projected onto the first $k$ principal components (i.e., eigenvectors with the largest eigenvalues) as defined in Equation~\ref{eq:dim-pca}, where \( \mathbf{W} \in \mathbb{R}^{d \times k} \) contains the top $k$ eigenvectors. The parameter $k$, arbitrarily defined by the practitioner, is proportional to the amount of variance preserved in the new representation.

\begin{equation}
    \label{eq:dim-pca}
    \mathbf{Z} = \mathbf{X} \mathbf{W}
\end{equation}

The resulting matrix $Z$ comprises a linear approximation of the original data represented by $X$, simplifying clustering by reducing noise and redundant dimensions.

\paragraph{Uniform Manifold Approximation and Projection (UMAP)}

UMAP is a non‑linear DR technique that captures complex, non‑linear structures in the data by preserving local and global relationships~\cite{mcinnes2018umap}. Unlike PCA, which assumes linearity, UMAP seeks to uncover the best low-dimensional representation by approximating the data's intrinsic manifold. The algorithm involves two main stages: graph construction and graph layout optimisation.

Graph construction is performed by modelling the data as a weighted k-neighbour graph to represent the local structure of the data. A probabilistic measure derived from the distance between points determines the graph's weights. In the graph layout optimisation stage, UMAP optimises a low-dimensional representation of the data using a force-directed graph layout. The optimisation minimises the cross-entropy between the high-dimensional and low-dimensional representations, effectively preserving the topology of the original dataset.

As a result, UMAP finds the best-curved surface (i.e., subspace) to represent the intra-data structure. By preserving local neighbourhoods and global data dispersion, UMAP (i) provides a precise representation of non‑linear relationships and (ii) enhances cluster separability in low-dimensional space.

\textcolor{black}{
\paragraph{Alternative DR control (PaCMAP).}
As a robustness check, we replaced UMAP with PaCMAP while keeping all other stages of the pipeline fixed (feature stack, PCA dimensionality, clustering configuration, and hyperparameters). 
For $\varphi_1$, using 170 embedding components and 100 neighbors followed by PCA (2 components) and K-means ($k=4$), we obtained a Silhouette score of 0.5645, a Davies–Bouldin index of 0.5971, and a Calinski–Harabasz index of 52583.1. 
These values are comparable to those obtained with UMAP, indicating that cluster separability is not materially affected by the specific choice of neighborhood-preserving DR method.
}

\paragraph{Combining UMAP and PCA}

In the proposed pipeline, DR is performed by connecting the UMAP output to the PCA input. UMAP is used to identify and capture non-linear patterns in the high-dimensional feature space, while PCA subsequently refines this reduction linearly. The parameters for both methods are fine-tuned to maximize the silhouette score, ensuring optimal cluster cohesion and separation. This approach leverages the strengths of both techniques, providing an effective representation of the data for clustering.

\textcolor{black}{
We apply UMAP before PCA (see Figure~\ref{fig:pipeline}) because UMAP is the main non-linear manifold learning step and benefits from operating directly on the full high-dimensional feature representation. PCA is then applied as a lightweight linear refinement/compression step to obtain a compact representation (2–3 dimensions) for visualization and to improve clustering stability. In contrast, applying PCA before UMAP would impose an a priori linear projection that may discard low-variance components that could still be relevant for defining neighbourhood relations in the original feature space. This design treats PCA not as a DR precursor, but as a final orthogonal projection of the UMAP embedding.
}

\subsubsection{Clustering}

The reduced representations are clustered using standard unsupervised algorithms (K-Means, Agglomerative Clustering, or Gaussian Mixture Models). No feedback
from the clustering stage is propagated upstream: feature extraction, DR, and clustering are strictly decoupled steps in the pipeline.

{\color{black}
The clustering problem addressed in this work is formulated and solved in a fully unsupervised setting. No physical labels are
assigned a priori to individual trajectories, and no dynamical diagnostic (e.g., resonance indicators, frequency analysis, or chaos 
indicators) is used during feature extraction, DR, or clustering.

In this context, standard external validation in the supervised learning sense, namely, comparison against a complete and predefined ground truth partition, is not directly applicable, as no such global labelling exists for the ensemble under study.

In Hamiltonian systems such as the one considered here, dynamical regimes including corotation, Lindblad resonances, and chaotic behaviour are well characterised locally in phase space and extensively documented in the literature. However, these regimes do not define a global, exhaustive, and mutually exclusive partition of all trajectories. Many orbits occupy transitional regions, exhibit mixed behaviour depending on the timescale considered, or require multiple diagnostic tools to be meaningfully characterised. Consequently, assigning a single, unambiguous dynamical label to each of the 22,288 orbits would require additional modelling assumptions and long-term integrations beyond the scope of the present study.

For this reason, clustering performance is primarily assessed using internal validation indices—namely the Silhouette score, the Davies–Bouldin index, and the Calinski–Harabasz index—which are appropriate in fully unsupervised scenarios and are used for model selection and pipeline comparison.
}

\textcolor{black}{
The number of clusters $k$ is not treated in this work as a purely data-driven hyperparameter, but rather as a physics-informed modelling choice. In Hamiltonian dynamical systems, the phase space is often structured into a small number of well-characterised regimes (e.g., libration zones, circulation domains, chaotic layers, and non-resonant regions), which motivates selecting $k$ according to the expected qualitative partition of the dynamics. In the simple pendulum validation test, $k=3$ was intentionally chosen to match the three canonical regimes visible in the Hamiltonian portrait (oscillation, prograde circulation, and retrograde circulation). In the Saturnian resonance dataset, we set $k=4$ to remain consistent with the dynamical mapping reported by Callegari and Yokoyama~\cite{callyoko2020}, who identify four dominant orbital regimes in the 11:10 Anthe–Mimas resonance region. This strategy ensures that the clustering output remains interpretable in terms of established dynamical behaviour rather than being solely driven by statistical criteria.
}

Physical knowledge of the system is incorporated a posteriori through the interpretation of the resulting clusters in dynamical maps and their qualitative consistency with known phase-space structures reported in previous studies. This provides a physics-informed interpretative validation complementary to the internal indices.


\subsubsection{Distance Metrics}

The choice of distance metrics plays a pivotal role in clustering algorithms, especially for time series data. In this study, the Euclidean distance was employed as the main metric due to its computational efficiency, mathematical simplicity, and compatibility with clustering methods such as K-Means and Gaussian Mixture Models. Euclidean distance remains the default measure for these methods, as they assume centroid-based optimisation in feature space \cite{bishop2006pattern}.

\paragraph{Dynamic Time Warping (DTW):} Dynamic Time Warping (DTW) is widely recognized for its ability to align time series sequences by handling temporal shifts and distortions \cite{berndt1994using}. However, DTW was not used in this study for the following reasons: \begin{itemize} \item \textbf{Computational Overhead:} Constructing a pairwise DTW distance matrix for a dataset of $22,288$ time series, each of length $400$, incurs a prohibitive time complexity of $\mathcal{O}(N^2 T^2)$, where $N$ is the number of series and $T$ is the series length \cite{sakoe1978dynamic}. \item \textbf{Suboptimal Results:} Initial experiments revealed no significant performance improvement with DTW compared to Euclidean distance, likely due to the extracted feature space reducing the need for temporal alignment. \end{itemize}

\paragraph{Alternative Metrics:} While Euclidean distance performed well in this study, other distance metrics remain relevant for time series clustering and high-dimensional feature spaces:
\begin{itemize}
    \item \textbf{Manhattan Distance (L1 Norm):} Robust to outliers, it computes the sum of absolute differences \cite{jain1999data}.
    \item \textbf{Cosine Distance:} Particularly useful in high-dimensional spaces where the angular similarity is more informative than magnitude differences \cite{tan2006introduction}.
    \item \textbf{Correlation Distance:} Suitable for clustering time series with similar trends but differing amplitudes \cite{shokoohi2015non}.
    \item \textbf{Mahalanobis Distance:} Effective for feature spaces with correlated variables and differing variances by incorporating the covariance matrix \cite{de2000mahalanobis}. \end{itemize}

\paragraph{Future Considerations:} Although DTW and other advanced distance metrics are beneficial for time series clustering, their computational cost must be balanced against dataset size and runtime constraints. Incorporating DR techniques (e.g., PCA, UMAP) prior to clustering can alleviate computational burdens while preserving structural relationships in the data \cite{van2008visualizing, mcinnes2018umap}.

\textcolor{black}{
Cosine or Mahalanobis distances can be beneficial in some high-dimensional settings. However, we adopted Euclidean distance primarily because (i) it is the native objective for centroid-based clustering and Gaussian mixtures, (ii) it remains stable and efficient at the scale considered, and (iii) the feature engineering stage (per-series z-normalization and heterogeneous descriptors) already mitigates scale effects. A systematic comparison of alternative metrics (e.g., cosine, whitening) is a natural extension and is left for future work.
}

\subsection{Outlier Repositioning via Graph Diffusion (ORG-D)} 

\begin{figure}[ht]
    \centering
    \includegraphics[width=0.85\textwidth]{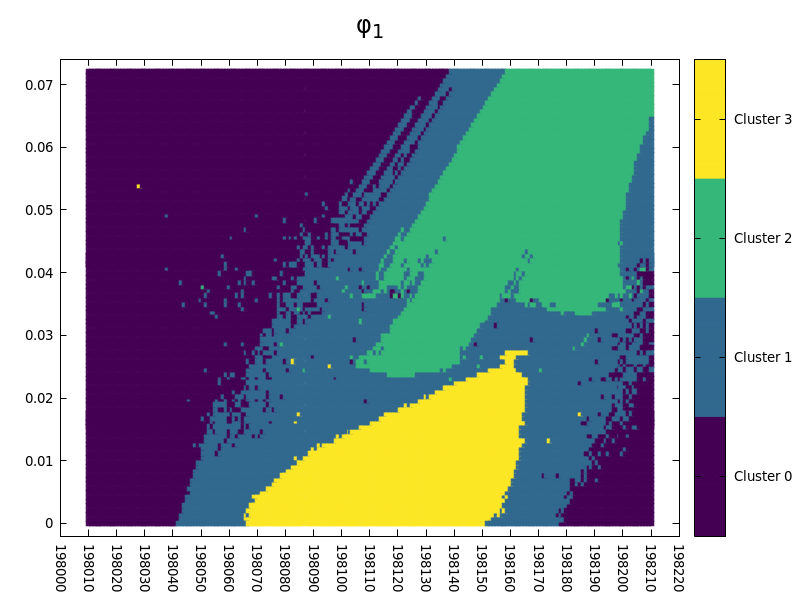}
    \caption{Dynamical map representing the clustering of initial orbital conditions in the space of semi-major axis versus initial eccentricity for the angle $\varphi_1$. The colour bar indicates different clusters obtained through the K-Means algorithm, capturing distinct dynamical behaviours in the orbital phase space.
    The horizontal axis represents the semi-major axis, and the vertical axis represents the eccentricity. Outliers are evident throughout the dynamical map.}
    \label{fig:dynamic_map_phi1}
\end{figure}

\begin{figure}[ht]
    \centering
    \includegraphics[width=0.85\textwidth]{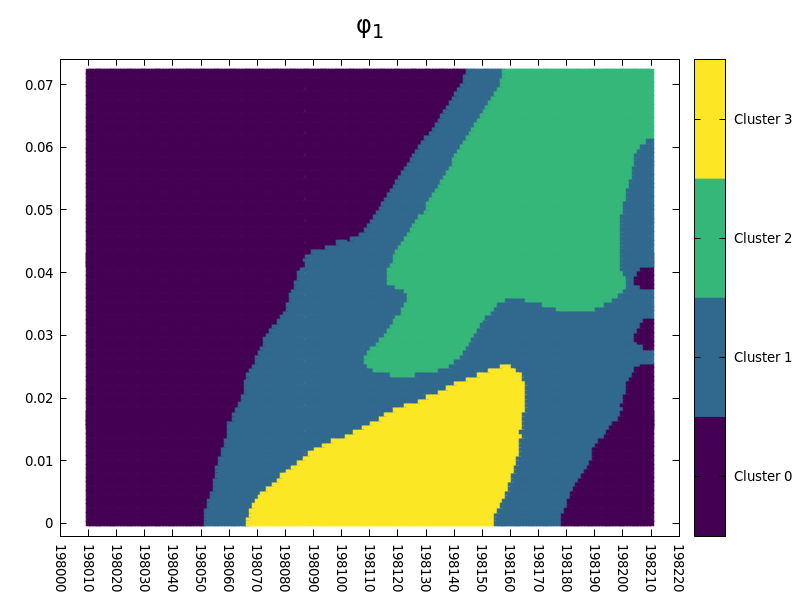}
    \caption{The same context of Figure~\ref{fig:dynamic_map_phi1} with outliers repositioned using $K=24$ nearest neighbours.}
    \label{fig:dynamic_map_phi1_K_24}
\end{figure}

\begin{figure}[ht]
    \centering
    \includegraphics[width=0.85\textwidth]{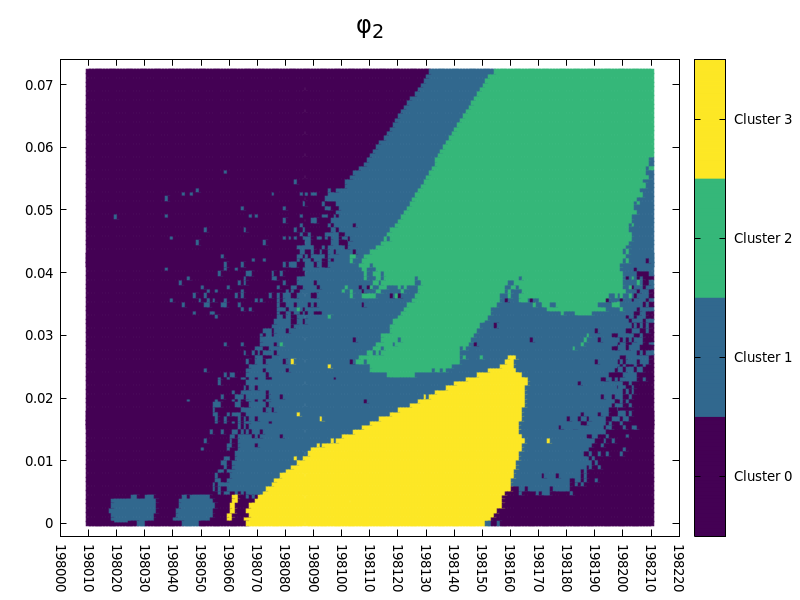}
    \caption{As in Figure~\ref{fig:dynamic_map_phi1} but for the angle $\varphi_2$.}
    \label{fig:dynamic_map_phi2}
\end{figure}

\begin{figure}[ht]
    \centering
    \includegraphics[width=0.85\textwidth]{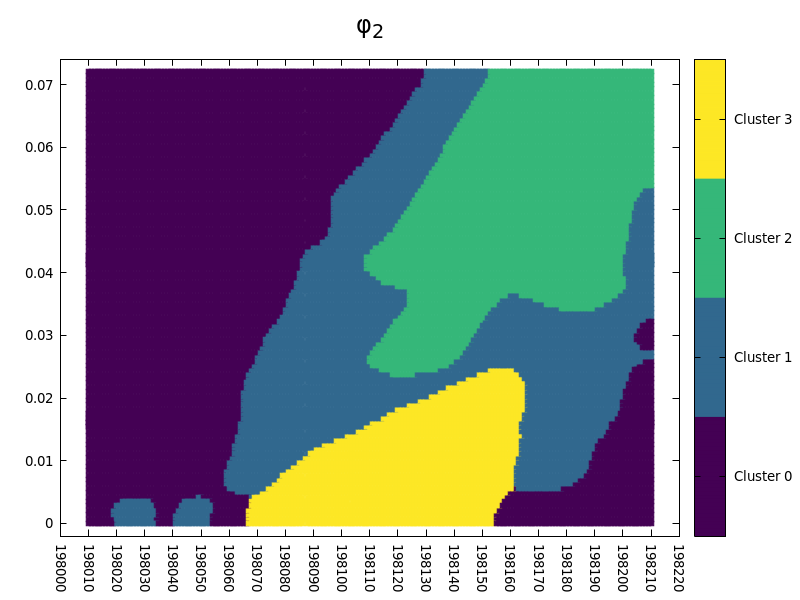}
    \caption{The same context of Figure~\ref{fig:dynamic_map_phi2} with outliers repositioned using $K=24$ nearest neighbours.}
    \label{fig:dynamic_map_phi2_K_24}
\end{figure}

The output of the clustering step often contains noise due to various factors; i.e., some elements of one cluster are mistakenly assigned to another. Particle Competition and Cooperation (PCC) \cite{breve2012particle} was originally a nature-inspired graph-based semi-supervised classification method. However, in this paper, it is used to eliminate noise by repatriating elements to their rightful cluster. This application is possible because the method was designed with the assumptions of cluster and label smoothness in mind. For instance, according to Figure~\ref{fig:dynamic_map_phi1}, there are some misplaced points in the wrong clusters. For example, the leftmost yellow point, labelled cluster 3, is located in a region typical of cluster 0.

First, a graph is built where each node represents a satellite orbit. Each node is connected to its $k$ nearest neighbours, based on their respective values of semi-major axis and eccentricity. The graph is then fed to PCC.

The nodes are randomly divided into 10 folds. Nodes belonging to one of the folds are presented to PCC with their respective cluster labels, while the remaining are presented unlabelled. The algorithm will spread the labels from the labelled nodes to the unlabelled nodes using a transductive approach, which works as follows.

A particle is generated for each labelled node, which will be called the home node of the particle. Particles will walk around the graph, trying to dominate as many nodes as possible. Particles that have the same label are said to belong to the same team and act cooperatively, while competing with particles from other labels/teams. It is an iterative process where particles use a random-greedy approach to decide which node to visit next, prioritizing nodes closer to their home node and nodes that their team has higher domination over.

At the end of the iterative process, the long-term domination levels \cite{breve2013fuzzy} are used as the probabilities for all labelled and unlabelled nodes. The whole process is repeated for each of the 10 folds. The entire fold division is repeated 100 times, for a total of 1000 PCC executions, so the probabilities for each node are the average of 1000 executions, maximizing their reliability.

After all executions, if an orbit has a higher probability of belonging to another cluster than the one to which it was assigned in the clustering step, it is repatriated, i.e., re-grouped to its rightful cluster.

Fig~\ref{fig:dynamic_map_phi1_K_24} shows the results of this process applied to the image in Fig~\ref{fig:dynamic_map_phi1} using $k=24$. Similarly, Fig~\ref{fig:dynamic_map_phi2_K_24} shows the results of this process applied to the image in Fig~\ref{fig:dynamic_map_phi2} using $k=24$. From the comparison of the two pairs of figures, it is clear that the outliers were effectively relabelled.

\begin{figure}[ht]
    \centering
    \includegraphics[width=0.88\textwidth]{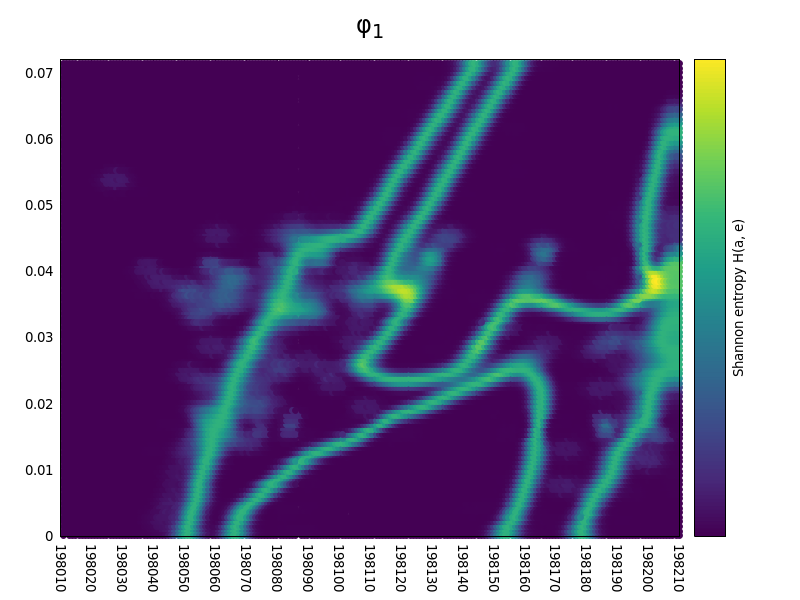}
    \caption{Entropy map in the $(a,e)$ plane for $\varphi_1$, computed from PCC-based soft labels using Eq.~\eqref{eq:entropy_tmp}. The field is shown as the 
    entropy per $(a,e)$ bin. 
    }
    \label{fig:entropy_map_phi1}
\end{figure}

\begin{figure}[ht]
    \centering
    \includegraphics[width=0.88\textwidth]{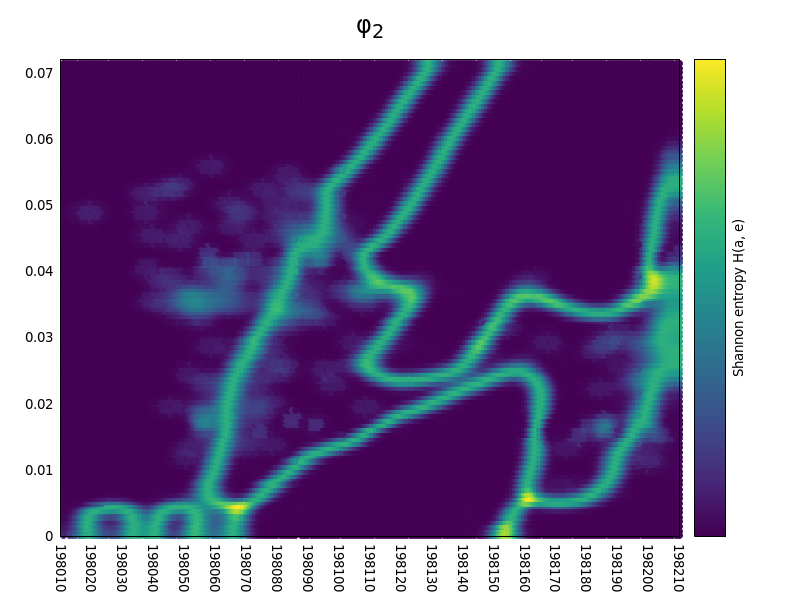}
    \caption{Entropy map in the $(a,e)$ plane for $\varphi_2$, computed from PCC-based soft labels using Eq.~\eqref{eq:entropy_tmp}. The field is shown as the
    entropy per $(a,e)$ bin. 
    }
    \label{fig:entropy_map_phi2}
\end{figure}

{\color{black}
\subsubsection{PCC-based soft labels and per-orbit entropy}\label{sec:pcc_uncertainty_tmp}

To quantify the confidence of the PCC-based repatriation step, we compute a \emph{per-orbit} uncertainty score from the PCC ``dominance levels''.
For each orbit (graph node) $i$, PCC yields a soft-label vector
$\mathbf{p}_i = (p_{i,1},\ldots,p_{i,K})$,
where $p_{i,k}\in[0,1]$ represents the (normalized) long-term domination level of team $k$ at node $i$, and $\sum_{k=1}^{K} p_{i,k}=1$.
In our implementation, the final $\mathbf{p}_i$ is obtained by averaging multiple PCC runs (different random fold assignments) and then renormalizing.

We then define the Shannon entropy of the soft labels as
\begin{equation}\label{eq:entropy_tmp}
H_i \;=\; -\sum_{k=1}^{K} p_{i,k}\,\log\!\big(p_{i,k}+\varepsilon\big),
\end{equation}
where $\varepsilon$ is a small constant used only for numerical stability.
Unless stated otherwise, $\log(\cdot)$ denotes the natural logarithm; using $\log_2$ changes $H_i$ only by a constant scaling factor.
Entropy is computed \emph{per orbit}: $H_i\approx 0$ indicates a highly dominant team (near-deterministic assignment), while larger values indicate ambiguous membership across teams.
The theoretical maximum is $H_i=\log K$ for the uniform distribution $p_{i,k}=1/K$.

For visualization in the $(a,e)$ plane, we define an entropy field $H(a,e)$ by binning orbits on a regular grid in $(a,e)$ and plotting the mean entropy per bin (this binning is used \emph{only} for visualization purposes).
Figures~\ref{fig:entropy_map_phi1}--\ref{fig:entropy_map_phi2} show entropy maps for $\varphi_1$ and $\varphi_2$.
Importantly, these maps do not aim to reproduce the categorical dynamical map; instead, they summarize the spatial distribution of \emph{soft-label uncertainty} induced by the PCC-based repatriation method.
}

\begin{table}[ht]
\centering
\caption{Summary of ORG-D impact on the dynamical maps for $\varphi_1$ and $\varphi_2$. 
For each angle we report the fraction of relabelled orbits and several agreement indices 
between pre-- and post--ORG-D cluster assignments.}
\label{tab:orm-metrics}
\begin{tabular}{lcccccc}
\hline
Angle & Relabelled & ARI & NMI & Homogeneity & Completeness & V-measure \\
\hline
$\varphi_1$ & $978$ (4.39\%) & 0.881 & 0.850 & 0.850 & 0.850 & 0.850 \\
$\varphi_2$ & $998$ (4.48\%) & 0.875 & 0.850 & 0.849 & 0.850 & 0.850 \\
\hline
\end{tabular}
\end{table}

\begin{table}[ht]
\centering
\caption{Confusion matrix between pre-- and post--ORG-D labels for $\varphi_1$.
Rows correspond to the original K-Means clusters and columns to the PCC-based ORG-D labels.}
\label{tab:orm-confusion-phi1}
\begin{tabular}{lcccc}
\hline
 & Post 0 & Post 1 & Post 2 & Post 3 \\
\hline
Pre 0 & 9315 & 367 & 3   & 0   \\
Pre 1 & 262  & 5222& 129 & 12  \\
Pre 2 & 4    & 137 & 4312& 0   \\
Pre 3 & 1    & 63  & 0   & 2461\\
\hline
\end{tabular}
\end{table}

\begin{table}[ht]
\centering
\caption{Confusion matrix between pre-- and post--ORG-D labels for $\varphi_2$.
Rows correspond to the original K-Means clusters and columns to the PCC-based ORG-D labels.}
\label{tab:orm-confusion-phi2}
\begin{tabular}{lcccc}
\hline
 & Post 0 & Post 1 & Post 2 & Post 3 \\
\hline
Pre 0 & 8804 & 369 & 1   & 9   \\
Pre 1 & 355  & 5248& 82  & 12  \\
Pre 2 & 0    & 112 & 4795& 0   \\
Pre 3 & 14   & 44  & 0   & 2443\\
\hline
\end{tabular}
\end{table}

Most of the trajectories reassigned by the PCC/ORG-D procedure are located in regions of elevated entropy (lighter lanes) in the PCC-induced soft labelling, consistently observed for both angular variables, $\varphi_1$ and $\varphi_2$. This indicates that the reassignment step primarily affects trajectories lying in locally ambiguous regions of the embedding space, while leaving well-defined regimes largely unchanged.

\subsubsection{Partition agreement measures (ARI, NMI)}
\label{sec:partition_agreement}

\textcolor{black}{
To quantify the global impact of the PCC-based ORG-D on the original clustering, we employ standard \emph{partition agreement measures}. These indices compare two partitions of the \emph{same dataset} -- the cluster assignments obtained before and after the ORG-D step -- and therefore do not rely on any external ground truth labels. Their purpose is not to assess clustering quality, but to measure how strongly the ORG-D modifies the initial partition structure.
}

Let $U=\{U_1,\dots,U_K\}$ and $V=\{V_1,\dots,V_K\}$ denote the partitions produced by
K-Means before and after the PCC-based repatriation, respectively.
\paragraph{Adjusted Rand Index (ARI).}
The Adjusted Rand Index (ARI) measures the similarity between two partitions while
correcting for chance agreement~\cite{hubert1985comparing}. It is defined as
\begin{equation}
\mathrm{ARI} =
\frac{\mathrm{RI} - \mathbb{E}[\mathrm{RI}]}{\max(\mathrm{RI}) - \mathbb{E}[\mathrm{RI}]},
\end{equation}
where $\mathrm{RI}$ denotes the Rand Index. ARI takes values in the interval
$[-1,1]$, with $\mathrm{ARI}=1$ indicating identical partitions and values close to
zero corresponding to random agreement.

\paragraph{Normalized Mutual Information (NMI).}
Normalized Mutual Information (NMI) quantifies the amount of shared information
between two partitions using information-theoretic concepts~\cite{strehl2002cluster}.
It is defined as
\begin{equation}
\mathrm{NMI}(U,V) =
\frac{2\,I(U,V)}{H(U)+H(V)},
\end{equation}
where $I(U,V)$ is the mutual information between partitions $U$ and $V$, and
$H(\cdot)$ denotes Shannon entropy. NMI ranges from $0$ (no shared information) to
$1$ (perfect agreement).

In this work, ARI and NMI are used to quantify the agreement between the
original K-Means clustering and the post-ORG-D partition. High values indicate
that ORG-D acts as a \emph{conservative refinement} of the initial clustering,
reassigning only a limited number of trajectories while preserving the global
structure of the partition.

To summarise the global impact of the PCC-based ORG-D on the dynamical maps,
Table~\ref{tab:orm-metrics} reports, for both critical angles $\varphi_1$ and
$\varphi_2$, the fraction of relabelled orbits together with the agreement indices
between pre-- and post--ORG-D cluster assignments (ARI, NMI, homogeneity,
completeness and V-measure). In both cases, only about $4.4\%$ of the 22\,288
trajectories are reassigned, while the agreement scores remain high (ARI
$\approx 0.88$, NMI and V-measure $\approx 0.85$). These values confirm that
ORG-D does not alter the global partition structure but rather performs targeted
corrections concentrated in dynamically ambiguous regions.

The detailed redistribution of orbits between clusters is shown by the confusion
matrices in Tables~\ref{tab:orm-confusion-phi1}
and~\ref{tab:orm-confusion-phi2}. Most entries lie on the diagonal, whereas the
dominant off-diagonal terms correspond to transitions such as
$0\rightarrow 1$ and $1\rightarrow 0$, concentrated near separatrix-like
boundaries between the non-physical and chaotic domains, consistently with the
high-entropy regions highlighted by the ORG-D uncertainty maps.

\subsection{Evaluation Metrics}
Two distinct classes of evaluation metrics are employed in this study, reflecting
different methodological stages of the proposed pipeline.

The first class comprises \emph{internal clustering validity indices}, namely the
silhouette score, the Davies--Bouldin (DB) index, and the Calinski--Harabasz (CH)
index. These metrics assess cluster compactness and separation directly from the
embedded feature space and do not rely on any external labels. They are therefore
used exclusively to compare different feature-extraction, dimensionality-reduction,
and clustering configurations during model selection.

The \textbf{silhouette score} \cite{rousseeuw1987silhouettes} $S$ measures cluster cohesion and separation. Equation~\ref{eq:silhouette} defines the computation of $S$, where \( a(i) \) is the average intra-cluster distance, and \( b(i) \) is the average nearest-cluster distance.

\begin{equation}
    \label{eq:silhouette}
    S = \frac{b(i) - a(i)}{\max(a(i), b(i))}
\end{equation}

The \textbf{DB} index \cite{davies1979cluster} assesses compactness and separation. It is computed as in Equation~\ref{eq:db}, where \( \sigma_i \) is the intra-cluster scatter of cluster \( i \).

\begin{equation}
    \label{eq:db}
    DB = \frac{1}{K} \sum_{i=1}^{K} \max_{j \neq i} \frac{\sigma_i + \sigma_j}{\|\mu_i - \mu_j\|}
\end{equation}

The \textbf{CH} index \cite{calinski1974dendrite} measures the ratio of dispersion between clusters to dispersion within clusters. It is computed as in Equation~\ref{eq:ch}, where \( \mathbf{B} \) and \( \mathbf{W} \) are the scatter matrices between clusters and within clusters.

\begin{equation}
    \label{eq:ch}
     CH = \frac{\text{tr}(\mathbf{B}) / (K - 1)}{\text{tr}(\mathbf{W}) / (n - K)}
\end{equation}

The second class comprises \emph{partition agreement measures}, which are employed
only to quantify the impact of the ORG-D based on
Particle Competition and Cooperation (PCC). In this case, the comparison is performed
between two partitions of the same dataset: the original clustering obtained by
K-Means and the refined clustering obtained after PCC-based relabelling.

These measures do not assume any ground truth physical labels. Instead, they quantify
the degree of consistency between pre-- and post--ORG-D assignments, allowing us to
assess whether the repatriation procedure preserves the global cluster structure.

\begin{table}[ht]
\centering
\caption{Controlled feature ablation under fixed DR and clustering hyperparameters for $\varphi_1$ and $\varphi_2$.}
\label{tab:ablation_both}
\begin{tabular}{llccc}
\hline
Angle & Feature Configuration & Silhouette & Davies--Bouldin & Calinski--Harabasz \\
\hline
$\varphi_1$ & MiniRocket & 0.5656 & 0.6593 & 42851.35 \\
$\varphi_1$ & MiniRocket + FFT & 0.5761 & 0.6466 & 43808.21 \\
$\varphi_1$ & MiniRocket + FFT + Wavelet & 0.5601 & 0.6695 & 35547.92 \\
$\varphi_1$ & Full Feature Stack & 0.5186 & 0.6979 & 33617.59 \\
\hline
$\varphi_2$ & MiniRocket & 0.6228 & 0.5108 & 89090.91 \\
$\varphi_2$ & MiniRocket + FFT & 0.6191 & 0.5155 & 89704.76 \\
$\varphi_2$ & MiniRocket + FFT + Wavelet & 0.6134 & 0.5129 & 76760.33 \\
$\varphi_2$ & Full Feature Stack & 0.6141 & 0.5062 & 78333.39 \\
\hline
\end{tabular}\label{tb7}
\end{table}

\begin{figure}[ht]
\centering
\includegraphics[width=0.8\textwidth]{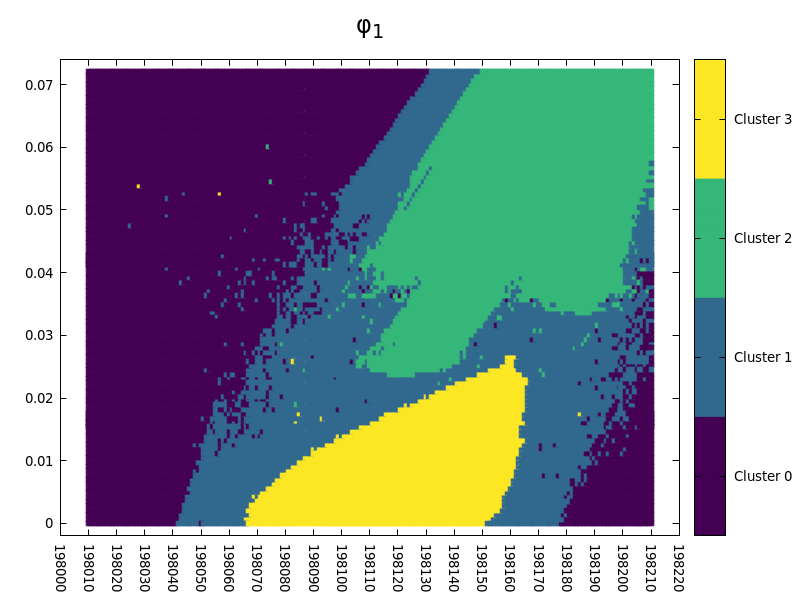}
\caption{Dynamical map obtained using the reversed DR ordering (PCA→UMAP) for $\varphi_1$. 
Although this configuration yields slightly higher internal validation indices, the chaotic transition region exhibits increased fragmentation and irregular cluster boundaries. 
This illustrates that improved geometric compactness in the embedding space does not necessarily correspond to enhanced dynamical coherence.}
\label{fig:pca_umap_phi1}
\end{figure}

\subsection{Feature Ablation and DR Sensitivity}

To address the referee’s request regarding feature sensitivity and DR ordering, we conducted controlled ablation experiments under frozen DR and clustering hyperparameters. 
In contrast to the optimisation procedure reported in Table~\ref{tab:compact_pipeline_benchmarks}, where hyperparameters were individually tuned for each configuration, the present analysis was performed with identical UMAP, PCA, and clustering settings across all feature combinations. 
This controlled protocol ensures that performance differences reflect the intrinsic contribution of the feature sets rather than secondary effects introduced by hyperparameter re-tuning.

We evaluated four representative feature configurations defined in the feature dictionary:
\begin{enumerate}
    \item MiniRocket features only,
    \item MiniRocket + FFT,
    \item MiniRocket + FFT + Wavelets,
    \item Full feature stack (MiniRocket + FFT + Wavelets + TSFresh).
\end{enumerate}

For each case, DR parameters were fixed, and clustering was performed using K-Means with identical settings across all runs. 
The resulting internal validation indices are reported in Table~\ref{tb7} for $\varphi_1$ and $\varphi_2$.

The ablation results indicate that MiniRocket features alone already capture a substantial portion of the dynamical structure. 
The addition of FFT descriptors improves cluster compactness and separation, as reflected by higher Silhouette and Calinski–Harabasz indices. 
The inclusion of wavelet and TSFresh features produces incremental but diminishing gains. 
Importantly, the large-scale dynamical partition remains qualitatively stable across configurations, indicating that the clustering is not driven by a single dominant feature family.

To further assess sensitivity to DR ordering, we evaluated the reversed configuration (PCA→UMAP). 
In this experiment, the output dimensionalities were correspondingly swapped to maintain comparable embedding scales (i.e., PCA was first applied with 30 components followed by UMAP with 3 components, whereas in the original configuration UMAP was applied with 30 components followed by PCA with 3 components). 
This adjustment preserves the final embedding dimensionality while isolating the effect of transformation order.

Although the reversed configuration yielded higher Silhouette values in certain cases, the corresponding dynamical map (Fig.~\ref{fig:pca_umap_phi1}) exhibits increased fragmentation within the chaotic transition region. 
In particular, the boundaries between resonant and chaotic domains become less coherent and more spatially dispersed. 
This behaviour indicates that improvements in geometric compactness within the embedding space do not necessarily correspond to improved dynamical consistency.

For this reason, the original UMAP→PCA ordering was retained in the main analysis, as it preserves the large-scale dynamical structures more faithfully.

It is worth emphasising that higher internal validation indices do not automatically imply physically more meaningful partitions in Hamiltonian dynamical systems. 
The Silhouette score measures geometric compactness in embedding space rather than dynamical coherence. 
Therefore, moderate numerical improvements under alternative DR orderings should not be interpreted as superior physical segmentation.

Overall, these experiments demonstrate that the proposed framework is robust with respect to feature composition and DR ordering, and that the principal dynamical regimes identified in this work remain stable under controlled perturbations of the pipeline.

\section{Fine Tuning}\label{sec:4}

The DR and clustering pipeline was fine-tuned using a grid search methodology. 
The process involved optimizing key parameters for both UMAP and PCA to maximize the silhouette score, ensuring optimal cluster cohesion and separation. This section outlines the fine-tuning procedure.

Initially, feature vectors were extracted from the dataset using various combinations of preprocessing methods. The feature combinations were systematically evaluated, with the most effective set determined based on clustering performance metrics. The explored feature combinations included FFT, wavelet, TSFresh, MiniRocket features, and their combinations. Each combination was assigned an identifier \( hkey \) and processed in batches to handle the dataset's size.

The fine-tuning process began with UMAP, where the following parameters were varied:
\begin{itemize}
    \item \textbf{n\_components}: The tested values ranged from 30 to 100, by intervals of 10.
    \item \textbf{n\_neighbours}: The tested values ranged from 30 to 100, by intervals of 10.
    \item \textbf{min\_dist}: The tested values were 0.0003125, 0.000625, and 0.00125.
\end{itemize}

UMAP was applied to reduce the dataset's dimensionality while preserving local and global structures. The output from UMAP was then further reduced using PCA to refine the dimensionality linearly. PCA was evaluated with parameters \textbf{n\_components} = 2, 3, 4, and 10.

For clustering, several methods were tested, including K-Means, Agglomerative Clustering, Gaussian Mixture Models (GMM), DBSCAN, and HDBSCAN. Each method underwent a grid search to identify the optimal hyperparameters for the clustering algorithm. For instance, K-Means was fine-tuned with variations in the initialization method, number of clusters, and maximum iterations.

The evaluation metrics for each configuration included:
\begin{itemize}
    \item \textbf{Silhouette Score}: A measure of cluster cohesion and separation.
    \item \textbf{Davies-Bouldin Index (DB)}: A metric for cluster compactness and separation.
    \item \textbf{Calinski-Harabasz Index (CH)}: The ratio of between-cluster dispersion to within-cluster dispersion.
\end{itemize}

These metrics were used to identify the best combination of feature vectors, UMAP parameters, PCA parameters, and clustering methods. This systematic approach ensured that the pipeline was optimised for both DR and clustering, yielding robust and meaningful clustering results for the dataset.

\textcolor{black}{
To address UMAP stochasticity and dimensionality-reduction stability, we fix the \texttt{random\_state} parameter in all UMAP executions, ensuring reproducible embeddings under the same configuration. UMAP hyperparameters were systematically explored via grid search, including \texttt{n\_neighbors} in the range $30–100$ and \texttt{min\_dist} in $\{0.0003125, 0.000625, 0.00125\}$, while varying the embedding dimensionality \texttt{(n\_components)} between $30$ and $100$. Since the goal of this study is to provide a deterministic and reproducible pipeline, we report results obtained under fixed-seed embeddings for the selected hyperparameters. 
}

\begin{figure}[htbp]
 \centering
 \includegraphics[width=9cm]{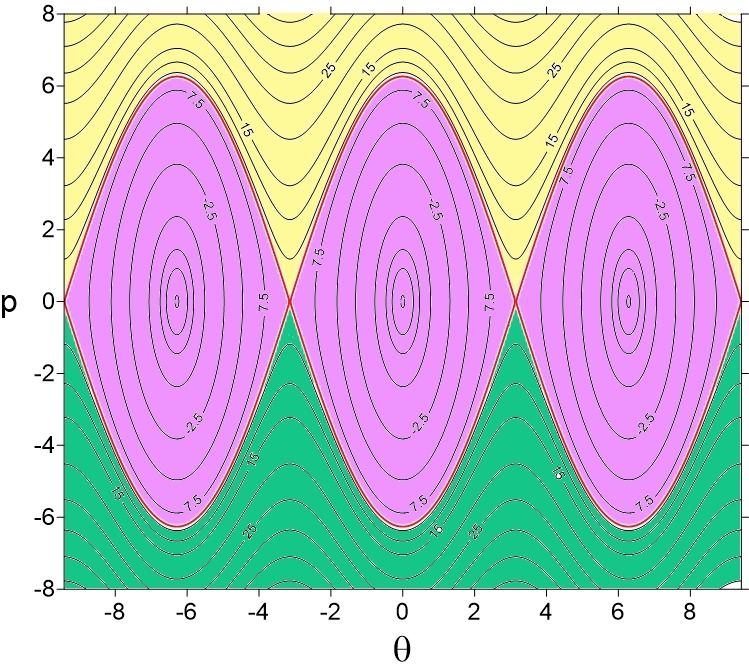}
  \caption{Level curves of the Hamiltonian (\ref{H}). Pink, green, and yellow indicate areas of the phase space corresponding to distinct regimes of motion: oscillation (pink), prograde circulation (yellow), and retrograde circulation (green). The red curves indicate the separatrix between the oscillatory and circulating regimes. Here, $m=l=1$ and $g=9.8$ (arbitrary units). }
  \label{F1}
\end{figure}

\begin{figure}[htbp]
 \centering
 \includegraphics[width=11cm]{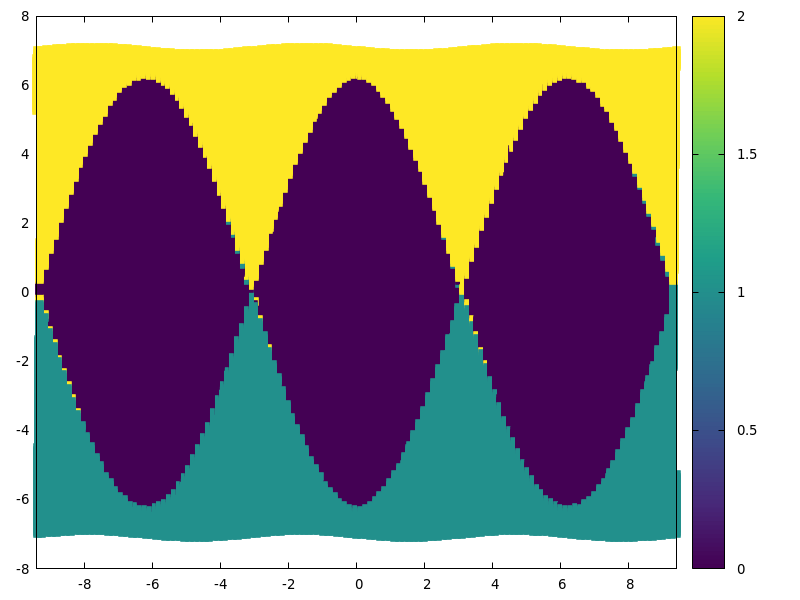}
  \caption{The resulting K-Means clustering, K = 3, for the simple pendulum time series, which is numerically computed in Figure~\ref{F1}}
    \label{F2}
\end{figure}

\section{Test: The dynamics of the simple pendulum}\label{sec:5}


A simple pendulum consists of a point mass $m$ suspended by a massless string of length $l$ from a fixed point $A$. Assuming that the motion is frictionless and confined to a vertical plane — and taking the pendulum’s lowest position (point $B$) as the reference level for potential energy — the Hamiltonian formulation of the problem is given by
\begin{equation}
H(\theta,p)=\frac{p^2}{2ml^2}+mgl(1-\cos\theta),\label{H}
\end{equation}
where $(\theta,p)$ is the pair of conjugate canonical variables. Here, $\theta$ is the angle between the string and the vertical line through $B$, and the conjugate momentum is defined as $p=ml^2\dot{\theta}$ (with $\dot{\theta}$ denoting the time derivative of $\theta$); $g$ represents the acceleration due to gravity.

The equations of motion are given by
\begin{eqnarray}
\dot{\theta}&=&\frac{\partial H}{\partial p}=\frac{p}{ml^2},\nonumber \\
\dot{p}&=&-\frac{\partial H}{\partial \theta}=-mgl\sin\theta.\label{EM}
\end{eqnarray}

It is well known that the analytical solution of the initial value problem (\ref{EM}) is cumbersome since it involves elliptic functions (e.g., \cite{FerrazMello2007}). For this reason, in many applications, numerical methods — such as Runge-Kutta algorithms — are often employed to obtain the trajectories (e.g., \cite{StoerBulirsch1993}). However, since the Hamiltonian has one degree of freedom, several aspects of the global dynamics of the pendulum can be inferred directly from its properties (\ref{H}).

Figure~\ref{F1} displays several level curves of the Hamiltonian in the representative phase space of the dynamical system, namely the plane $(\theta,p)$. In each of the three coloured regions in Figure~\ref{F1}, the motion of the pendulum is characterized by a specific type of behaviour: prograde circulation (yellow), retrograde circulation (green), and oscillations about the equilibrium point (pink level curves). The circulating regimes are separated from the oscillatory motion by the separatrix, which is indicated by the red curves\footnote{The motion of the pendulum starting from initial conditions corresponding to the Hamiltonian value at the separatrix ($H=9.799804688$ in Figure~\ref{F1}) tends asymptotically to the unstable equilibrium points located at $n\pi$, where $n\in\mathbb{Z}$.}

Applying the clustering pipeline shown in the main text, we have the following dynamical map shown in Figure~\ref{F2}, where the number of clusters, $K=3$, was intentionally chosen to match the number of equilibrium states depicted in Figure~\ref{F1}.

\begin{figure}[htbp]
    \centering
    \includegraphics[width=15cm]{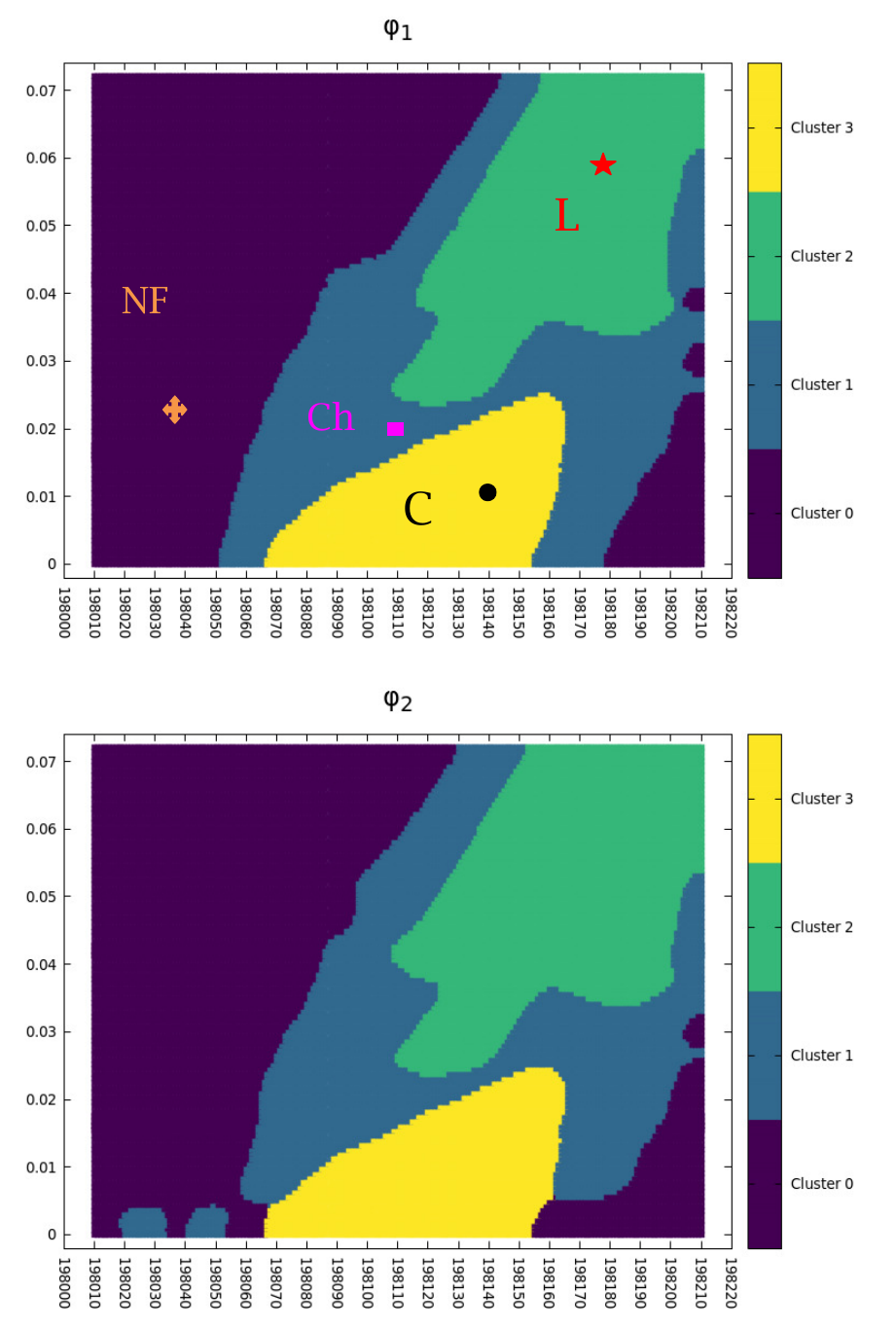}
    \caption{Dynamical maps representing the clustering of initial orbital conditions in the space of semi-major axis versus initial eccentricity. The top and bottom panels show the mapping for \(\varphi_1\) and \(\varphi_2\), respectively. The colour-coded regions indicate different clusters obtained through the K-Means algorithm, capturing distinct dynamical behaviours indicated by C, L, Ch, and NP, meaning corotation resonance, Lindblad resonance, chaotic motion, and non-physical meaning, respectively. At the top panel, four coloured symbols of the initial conditions of representative temporal series of each region are shown (see Figure~\ref{fig3}).}
    \label{fig2}
\end{figure}

\begin{figure}[htbp]
    \centering
    \includegraphics[width=1.\textwidth]{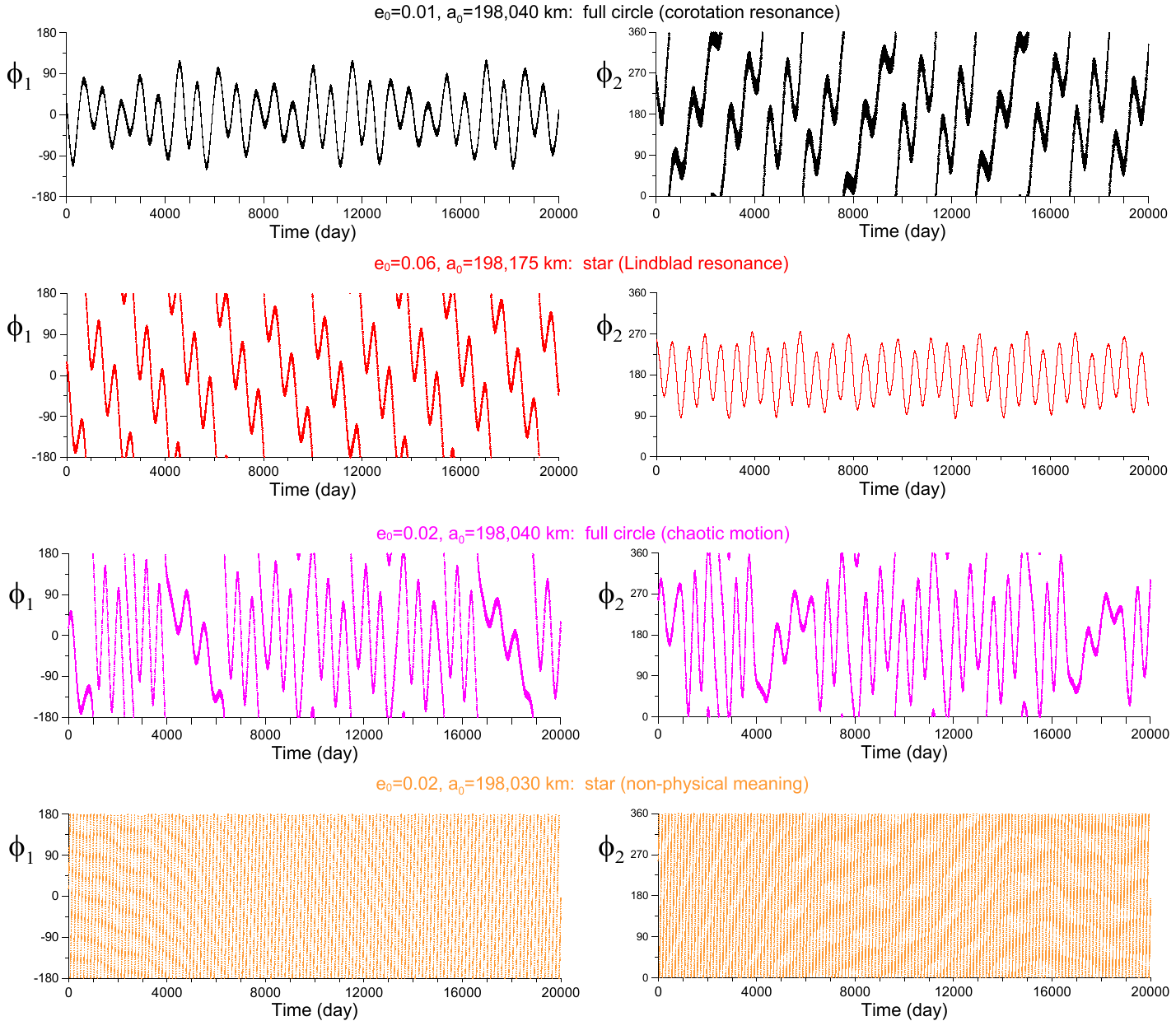}\caption{Time variations of the critical angles \(\varphi_1\) and \(\varphi_2\) corresponding to the initial conditions indicated the symbols given in Figure \ref{fig2}. From top to bottom: full circle, star, rectangle, crux. The corresponding initial values of the semi-major axis and the eccentricity of each orbit are given at the top of the panels.}
    \label{fig3}
\end{figure}

\section{Application: cluster analysis of simulations of clones of natural Saturn satellites}\label{sec:6}

This section presents the application of the methods described above. As noted in Section~\ref{sec:3}, the dataset analysed here was generated by numerical simulations of clones of natural satellites, incorporating the gravitational effects of Saturn’s zonal harmonics and major satellites. It comprises two variables — the corotation and Lindblad resonant angles\footnote{They are defined in terms of elliptic orbital elements as $\varphi_1 = 11 \lambda_S - 10 \lambda_M - \bar{\omega}_M$ and $\varphi_2 = 11 \lambda_S - 10 \lambda_M - \bar{\omega}_S$, where $\lambda_S$ and $\lambda_M$ are the mean longitudes of a test particle and the satellite Mimas, and $\bar{\omega}_M$ and $\bar{\omega}_S$ are the longitudes of pericentre of Mimas and of the test satellite, respectively~\cite{callyoko2020}.}, which are representative variables whose time variations provide informative descriptors of the dynamics of the particle ensemble.

Figure~\ref{fig2} shows two dynamical maps resulting from the techniques proposed in this work. The top (bottom) plot results from the time series analyses of the $\varphi_1$ ($\varphi_2$) angles. The methods were able to detect at least four very well-defined regions in the phase space of the semi-major axis versus the initial eccentricity of the test particles\footnote{\textcolor{black}{The choice of $k=4$ is justified after inspection of results given in Figure 7 in the paper {Callegari Jr.} and Yokoyama \cite{callyoko2020}, where the main regimes of motion of particles around the 11:10 resonance with the satellite Mimas have been mapped.}}. These regions are denoted, respectively, by Corotation resonance (yellow - cluster 3), Lindblad resonance (green - cluster 2), chaotic motion (blue - cluster 1), and non-physical meaning (purple - cluster 0). In order to illustrate the nature of these dynamical states of the particles, Figure~\ref{fig3} shows the time variations of $\varphi_1$ and $\varphi_2$ for four initial conditions within the domains of the main regions. The Corotation resonance is characterized by the perpetual oscillation of the angle $\varphi_1$ around zero while $\varphi_2$ circulates. The Lindblad resonance is characterized by the oscillation of the angle $\varphi_2$ around $\pi$ while $\varphi_1$ circulates. \textcolor{black}{In the region designated as chaotic, both angles alternate their regimes between oscillation and circulation in their respective ranges}. The initial conditions far from resonance domains define areas without any physical interest since no signature of the resonances appears. \textcolor{black}{In these cases, the critical angles suffer short-period time variations compared to the typical time-scale of resonant signatures.}

It is noteworthy that the application of the K-Means algorithm with short time series with 400 timesteps analysed in this work was able to reproduce the results of the dynamical maps given originally in Callegari \& Yokoyama's work, where huge time series have been utilised. In fact, a comparison of Figure~\ref{fig2} with the mapping of the 11:10 Anthe-Mimas resonance shown in Figure~\ref{fig3} in \cite{callyoko2020} gives good agreement with Figure~\ref{fig2}.

\textcolor{black}{
In terms of computational cost, to provide actionable guidance without overstating hardware-dependent benchmarks, we performed an indicative profiling run for the representative case of 22,288 × 400 time series, using the same software stack as in the main experiments. The measurements were obtained on a server-class machine running an Intel Xeon E5-2698 v4 with 80 cores and 1 TB of RAM; the pipeline is entirely CPU-based and does not rely on GPU acceleration.
}

In this configuration, the end-to-end pipeline required on the order of 10 minutes of wall-clock time per run, with a peak resident memory (RSS) of approximately 6.8 GB, which includes all threads (threads share the same address space). Feature extraction dominated the runtime, while UMAP and k-means contributed a smaller fraction of the total cost. As expected, absolute timings vary with CPU model, thread parallelism, and I/O; therefore, these values should be interpreted as order-of-magnitude guidance, while the relative cost across pipeline stages is robust.

\section{Conclusive Remarks}
\label{sec:7}

This work presented a fully reproducible pipeline for clustering large collections of astronomical orbital time series. We combined complementary feature extractors (MiniRocket, FFT, DWT, and TSFresh), non-linear dimensionality reduction (UMAP), and classical clustering (e.g., K-Means), and assessed the results with Silhouette, Davies--Bouldin, and Calinski--Harabasz indices. The resulting embeddings separate major dynamical regimes in a way that is consistent with domain knowledge, providing compact visual summaries that facilitate the inspection of resonant structures and transition regions.

A practical contribution of this study is a simple \emph{Outlier Repositioning via Graph Diffusion} (ORG-D) for placing rare or out-of-distribution (OOD) time series into an \emph{already learned} low-dimensional manifold \emph{without} distorting its geometry. The method decouples \emph{detection} from \emph{placement}: (i) learn the embedding on an in-distribution set only; (ii) flag potential outliers in the original feature space using a robust distance to the nearest cluster centre (e.g., median/MAD-normalized or Mahalanobis) and a high-quantile threshold; (iii) obtain two-dimensional coordinates for flagged items by applying the frozen UMAP transform and computing their position via $k$-NN barycentric interpolation over inlier neighbours; and (iv) render them with an uncertainty score (e.g., inverse local density) while avoiding any re-optimisation of the embedding. In practice, this keeps the global cluster layout stable and prevents outliers from pulling dense regions apart.

In our experiments, ORG-D consistently repositioned rare orbits at the periphery of the closest cluster or along separatrix-like boundaries, making them visually salient without altering the centroids or neighbourhood relations of in-distribution data. Computationally, once features are extracted, neighbour search and interpolation dominate the cost and scale sub-quadratically with dataset size when using standard approximate nearest-neighbour (ANN) indices.

The overall quality remains sensitive to hyperparameters (UMAP neighbourhood size, interpolation $k$, and the outlier threshold) and to the choice of base features. Dynamic Time Warping (DTW) was considered but not included due to its $\mathcal{O}(N^2 T^2)$ runtime on long series and because feature-space alternatives preserved temporal information sufficiently well for clustering at this scale.

A further contribution of this study is to consolidate and extend the application of clustering techniques within celestial mechanics. Earlier work in this area demonstrated that unsupervised learning can be used to classify orbital behaviours, separate resonance zones and the chaotic layers that surround them. Such an approach provides a complementary perspective to analytical and perturbative methods by revealing families of orbits and transitional dynamics that are otherwise difficult to capture. By embedding these ideas into a broader machine learning pipeline that integrates modern feature extraction, dimensionality reduction and validation indices, the present work strengthens the role of clustering as a rigorous tool for exploring large ensembles of resonant and chaotic dynamical systems.

Furthermore, starting from short, raw time series (400 samples) of the resonant angles, our pipeline qualitatively reproduces the phase-space maps that are traditionally obtained only after very long and computationally expensive integrations, recovering the main resonance domains and transitional regions (cf. Figure~\ref{fig2}).

Future improvements include: (a) benchmarking ORG-D with alternative detectors (e.g., Isolation Forest) and calibration strategies for uncertainty; (b) streaming/incremental embeddings to accommodate continuously generated orbits; and (c) physics-informed features that better capture libration/rotation switching and other resonance-driven phenomena.

\textcolor{black}
{Incremental and streaming extensions of the proposed framework constitute a natural direction for future work. In particular, an important practical scenario is the continuous generation of new orbital trajectories, which requires integrating out-of-sample time series into an already learned low-dimensional manifold without distorting its geometry.  A systematic evaluation of such a streaming setup would require dedicated experimental protocols to assess embedding stability, cluster consistency, and robustness under potential distribution shifts, and is, therefore, left for future investigation.
}

\ifblind
\else

\section*{Acknowledgements}

NCJ thanks São Paulo Research Foundation (FAPESP) for funding projects 2020/06807-7 and 2025/02325-1.

\fi

\section*{Data availability}
The datasets and full source code required to reproduce the experiments are publicly available at \url{https://github.com/epmarinho/ts-saturn-orbits}.

\begin{appendices}

\section{Clustering algorithms}\label{sec:distance-clustering}

Three algorithms were considered in this study: K-Means, agglomerative clustering, and Gaussian Mixture Models (GMM) clustering. This section briefly describes each of them.

{K-Means} partitions data into \( K \) clusters by minimizing within-cluster variance $\mathcal{V}$ defined in Equation~\ref{eq:k-means}, where \( \mu_k \) is the centroid of cluster \( C_k \) \cite{MacQueen1967}.

\begin{equation}
    \label{eq:k-means}
    \mathcal{V} = \sum_{k=1}^{K} \sum_{\mathbf{z} \in C_k} \|\mathbf{z} - \mathbf{\mu}_k\|^2
\end{equation}

{Agglomerative clustering} builds a hierarchy of clusters using linkage criteria such as the Ward’s method~\cite{ward1963hierarchical}, which computes the distance $d_\text{ward}(C_i,C_j)$ between clusters $C_i$ and $C_j$ based on Equation~\ref{eq:ward}, where \( C_i \) and \( C_j \) are clusters, and \( \mu_i \), \( \mu_j \) are their centroids.

\begin{equation}
    \label{eq:ward}
    d_\text{ward}(C_i,C_j) = \frac{|C_i||C_j|}{|C_i| + |C_j|} \|\mathbf{\mu}_i - \mathbf{\mu}_j\|^2
\end{equation}

 {Gaussian Mixture Models (GMM)} transform data using a mixture of Gaussians, optimising the log-likelihood $\mathcal{L}$ defined in Equation~\ref{eq:log-likelihood}, where \( \pi_k \), \( \mathbf{\mu}_k \), and \( \sigma_k \) are the weight, mean, and covariance of the \( k \)-th Gaussian, respectively \cite{bishop2006pattern}.

\begin{equation}
    \label{eq:log-likelihood}
    \mathcal{L} = \sum_{i=1}^{n} \log \left( \sum_{k=1}^{K} \pi_k \mathcal{N}(\mathbf{z}_i | \mu_k, \sigma_k) \right)
\end{equation}

Experiments were performed using each of these clustering algorithms, and the results were assessed according to a set of evaluation metrics. These metrics were employed to allow for comparing the results between different combinations of the techniques presented so far.





\end{appendices}



\end{document}